\documentclass[%
 reprint,
 amsmath,amssymb,
 aps,
 prx,
]{revtex4-2}

\usepackage{graphicx}
\usepackage{dcolumn}
\usepackage{bm}
\usepackage{color}

\usepackage[dvipdfmx]{}
\usepackage{comment}
\begin{document}

\newcommand{\pdif}{\partial}
\newcommand{\dif}{\mathrm{d}}
\newcommand{\Tpar}[3]{\left(\frac{\pdif #1}{\pdif #2} \right)_{#3}}
\newcommand{\TTpar}[3]{\left(\frac{\pdif^2 #1}{\pdif #2^2} \right)_{#3}}
\newcommand{\Par}[2]{\frac{\pdif #1}{\pdif #2}}
\newcommand{\Dif}[2]{\frac{\dif #1}{\dif #2}}
\newcommand{\PPar}[2]{\frac{\pdif^2 #1}{\pdif #2^2}}
\newcommand{\DDif}[2]{\frac{\dif^2 #1}{\dif #2^2}}
\newcommand{\ParPar}[3]{\frac{\pdif^2 #1}{\pdif #2 \pdif #3}}
\newcommand{\DifDif}[3]{\frac{\dif^2 #1}{\dif #2 \dif #3}}
\newcommand{\Totalpar}[3]{\Tpar{#1}{#2}{#3} \dif #2+\Tpar{#1}{#3}{#2} \dif #3}
\newcommand{\Mean}[1]{\langle #1 \rangle}


\title{Precipitation induced filament pattern of injected fluid controlled by structured cell}
\author{Shunsuke Tanaka}
\affiliation{%
 Department of Applied Physics, Tokyo University of Science, 6-3-1 Nijuku, Katsushika-ku, Tokyo, 125-8585, Japan}

\author{Kojiro Otoguro}
\affiliation{%
 Department of Applied Physics, Tokyo University of Science, 6-3-1 Nijuku, Katsushika-ku, Tokyo, 125-8585, Japan}

\author{Miyuki Kunihiro}
\affiliation{%
 Department of Applied Physics, Tokyo University of Science, 6-3-1 Nijuku, Katsushika-ku, Tokyo, 125-8585, Japan}
 
\author{Hiroki Ishikawa}
\affiliation{%
 Department of Applied Physics, Tokyo University of Science, 6-3-1 Nijuku, Katsushika-ku, Tokyo, 125-8585, Japan}
 
\author{Yutaka Sumino}
\email{ysumino@rs.tus.ac.jp}
\affiliation{%
 Department of Applied Physics, Tokyo University of Science, 6-3-1 Nijuku, Katsushika-ku, Tokyo, 125-8585, Japan}
 \affiliation{
Water Frontier Science \& Technology Research Center, and Division of Colloid Interface, Research Institute for Science \& Technology, Tokyo University of Science, 6-3-1 Nijuku, Katsushika-ku, Tokyo, 125-8585, Japan
}

\date{\today}

\begin{abstract}
Mixing of two fluids can lead to the formation of a precipitate. If one of the fluids is injected into a confined space filled with the other, a created precipitate disrupts the flow locally and forms complex spatiotemporal patterns. 
The relevance of controlling these patterns has been highlighted in the engineering and geological contexts. 
Here, we show that such injection patterns can be controlled consistently by injection rate and obstacles. Our experimental results revealed filament patterns for high injection and low reaction rates, and the injection rate can control the number of active filaments. Furthermore, appropriately spaced obstacles in the cells can straighten the motion of the advancing tip of the filament. A mathematical model based on a moving boundary adopting the effect of precipitation reproduced the phase diagram and the straight motion of filaments in structured cells. 
Our study clarifies the impact of the nonlinear permeability response on the precipitate density and that of the obstacles in the surrounding medium on the motion of the injected fluid with precipitation.
\end{abstract}

\maketitle

\section{Introduction}
Injection of fluid is long known to cause far-from-equilibrium pattern formation, such as viscous fingering, owing to the viscosity contrast of the injected and the displaced fluids~\cite{Homsy1987, Al-Housseiny2012}. Further complexity is introduced when a chemical reaction is involved, a reaction-injection system. Reaction-injection system create patterns owing to the dissolution of matrix~\cite{Liu2017}, the contrast of density~\cite{Haudin2014a}, the change in interfacial tension~\cite{Shi2006}, and the precipitation~\cite{Riolfo2012, Nagatsu2014}. Among these reaction-injection systems, the patterns caused by the precipitation have further relevance in fluid motion underground. The decrease in fluid permeability owing to the precipitation often creates filament patterns that counter-intuitively enhance fluid migration. The typical examples include CO$_2$ sequestration techniques~\cite{Berg2012a}, enhanced oil recovery~\cite{Thomas2008, Muggeridge2014} and chemical grouting~\cite{Karol2003}. We can also find geological examples related to the patterns caused by precipitation. These examples include an injection-induced seismology~\cite{Ellsworth2013, Guglielmi2015, Kim2018, Schultz2020} and the coupling of fluid migration and silica precipitate observed in the basic process of the regular earthquakes~\cite{Audet2014a, Otsubo2019}. All examples of the injection process shown here include the coupling between the flows and the precipitate, which reduces the system's permeability. 

\begin{figure}
   \centering
    \includegraphics[width=0.48\textwidth]{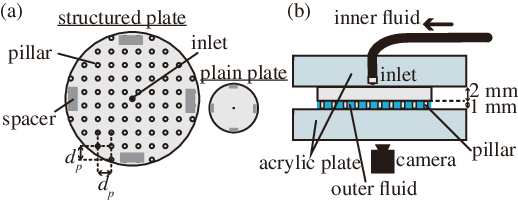}
    \caption{(a) Setup of top plates. Both structured cell and plain cell have a 70 mm in radius. A structured cell has pillars whose height and diameter are 1 mm. (b) Side view of the experimental setup. Inner fluid was injected from the inlet fixed at the center of the top plate. The fluid can escape from the gap between the plates without constraints. }
    \label{Fig:1}
\end{figure}

The behavior of fluid underground, thus, provides us with relevant information to predict the outcome of injection efficiently and accurately; however, observing fluid {\it in situ} is challenging. In this context, an experimental model system, a reaction-injection system with precipitation, is necessary. Previous model system~\cite{Haudin2014, Haudin2015a, Haudin2015b, Wagatsuma2017}, in which a metal salt solution is injected into a cell with a narrow gap, a Hele-Shaw cell, filled with an aqueous solution of silicate, find the appearance of anisotropic filament patterns which causes enhanced fluid migration to the remote position. Such filament patterns are universal ones observed independent of the chemical details~\cite{Podgorski2007, Nagatsu2008, Nagatsu2014,Niroobakhsh2017}. Care should be taken that a realistic geometry is not homogeneous because of the granular nature of soils. Therefore, the effect of spatial obstacles on the filament patterns should be investigated. Indeed, several trials have been conducted with non-planar cells to reveal the impact of the inhomogeneity of the system~\cite{Wang2020b}. However, owing to the explored parameter space, little effect was paid to a filament pattern that substantially impacts the transportation of the injected fluid. In this study, we conducted an injection experiment represented by Figs.~\ref{Fig:1} (a) and (b), using the micellar solution system and introduced inhomogeneity to a cell by fabricating regularly arrayed pillars at the cell's surface.

\begin{figure*}
   \centering
    \includegraphics[width=0.96\textwidth]{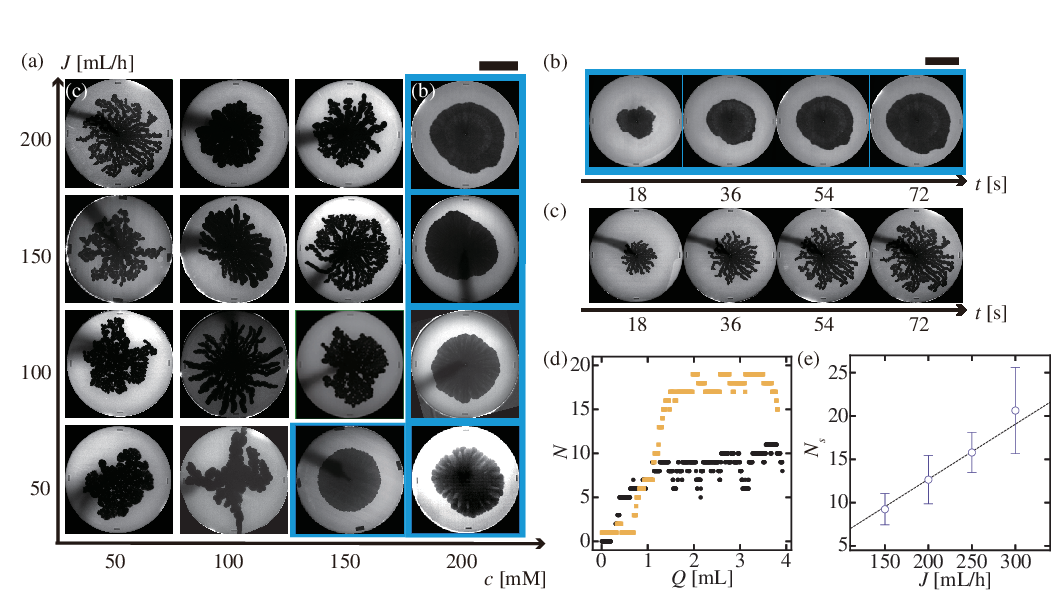}
    \caption{(a) Phase diagram of injection pattern for different $c$ and $J$ in a plain cell. Blue-shaded/no-shaded regions: circular/filament patterns. The time development of (b) circular ($c$ = 200 mM, $J$ = 200 mL/h) (c) filament ($c$ = 50 mM, $J$ = 200 mL/h) pattern. Scale bar: 50 mm. (d) The time course of the number of active filaments $N$ ($c$ = 50 mM). Black circles and yellow squares represent $J$=150 and 300 mL/h. (e) The saturated number of active filaments $N_s$ for $c$ = 50 mM. The error bar corresponds to the standard deviation. The dashed line represents the linear fitting.}
    \label{Fig:2}
\end{figure*}

\section{Experimental setup}
Water was purified using a Millipore Milli-Q system. Cetyl trimethyl ammonium bromide (CTAB) was purchased from Tokyo Chemical Industry Co., Ltd. Sodium salicylate (NaSal) and brilliant blue were purchased from Wako Pure Chemical Industries, Ltd. As the outer fluid, we used an aqueous solution of CTAB, whose concentration was maintained at 50 mM throughout the experiments. The injected inner fluid was an aqueous solution of NaSal stained with brilliant blue for visualization. We varied the concentration of NaSal as a parameter and denoted it as $c$. The amount of brilliant blue used in the experiment was less than 0.1 wt\%.

The experimental system consisted of a horizontal Hele-Shaw cell with a gap width of 1 mm, which is shown in Figs.~\ref{Fig:1} (a) and (b). The top plate of the cell was made of nylon and printed using a 3D printer service (DMM.make). The thickness and radius of the top plate were 2 mm and 70 mm, respectively. A plain cell was fabricated using a flat-top plate with spacers whose height was 1 mm. A structured cell was fabricated using a top plate with circular pillars with a height and diameter of 1 mm and spacers. These pillars were arranged in a square lattice, where the lattice constant, $d_p$, was varied as a parameter. The top plate was supported by two transparent acrylic plates with thicknesses of 5 mm. These plates are sufficiently thick to avoid cell deformation and maintain a constant injection rate of the inner fluid. A hole was bored at the center of the top plate, and inner fluid was injected from the hole. An outer fluid is initially introduced to fill the cell. The fluid escapes from the cell through the side part of the cell when the inner fluid is injected. 

The inner fluid was injected from the center of the top plate at various injection rates ($J$ = 50 to 300 mL/h). The central hole and a 10 mL syringe (Top Co. Ltd) were connected to a nylon tube (Nihon Pisco Co., Ltd.; internal diameter, 2.5 mm) whose length was 500 mm. Here, special care was taken to maintain a constant injection rate; that is, the selected nylon tube was sufficiently rigid. The syringe was installed on a syringe pump (CXF1010; ISIS Co. Ltd.). The system was illuminated from the top, and the pattern of injected fluid was measured from the bottom of the cell using a digital video camera at 10 Hz and analyzed using Image J software (NIH)~\cite{Schneider2012}.

\section{Results and Discussion}

When the inner and outer fluids are mixed, a gel made of worm-like micelles is formed. This gel behaves as an ideal Maxwellian fluid with a single relaxation time~\cite{Shikata1987, Shikata1989}.  Note that the viscosity of the inner and outer fluid is close to that of water in the absence of worm-like micelles (inner fluid $c$=50 mM, 1.03 m Pa s, outer fluid: 1.07 mPa s). Therefore, viscous fingering did not occur in the absence of the reaction.

\subsection{Plain cells}

Figure~\ref{Fig:2} (a) shows the situation after the injected fluid $Q$ reached 4 mL. Circular patterns were observed at a high $c$ and low $J$. The front of the inner fluid advanced in a radially symmetric manner while producing the layer of gel precipitate, as shown in Fig.~\ref{Fig:2} (b). However, we observed a filament pattern, where the front of the inner fluid becomes a 2-dimensional tube at low $c$ and high $J$. The filaments exhibited meandering during the extension, and some of these filaments halted their extension. It should be noted that this meandering motion was typical of the filament pattern observed in previous studies~\cite{Haudin2014, Podgorski2007, Wagatsuma2017}. Furthermore, new filaments were generated during the injection process by fracturing the layer of precipitate in a branching manner, as shown in Fig.~\ref{Fig:2} (c).



As reported in the previous studies~\cite{Podgorski2007, Wagatsuma2017}, the number of extending and active filaments reached a steady value during the injection. Such a saturation in the number of active filaments was revealed by image analysis. The motion of the front was obtained from the difference of images (D-images) with 2 s time intervals. Only the active filaments were visualized as the moving domain. The domain in D-images larger than 0.7 mm$^2$ is recognized as the front of the active filaments. The time course of the number of active filaments $N$ is plotted for $J$=150 (black circles) and 300 mL/h (yellow squares) in Fig.~\ref{Fig:2} (d). Here, we plotted $N$ against the amount of injected fluid, $Q=Jt$. Initially, the number of active filaments increased with $Q$; however, it reached a maximum steady value. We obtained the saturated number $N_s$ from the data averaged over $Q$=1.5-3.5 mL, where the initial transient behavior can be ignored. The relationship between $J$ and $N_s$ is illustrated in Fig.~\ref{Fig:2} (e). A monotonic increase in $N_s$ with $J$ was observed. The dashed line corresponds to the fitting with $N_s= \alpha J$, whereas $\alpha=6.35 \times 10^{-2}$ $\pm 1.3 \times 10^{-3}$ h/mL. We also confirmed that the speed of the actively extending filament was almost constant with $J$. Such a linear dependence of $N_s$ on $J$ and a constant front speed implies that the characteristic width of the filament is fixed and determined independently from $J$. These results are consistent and much improved in resolution compared with the previous study~\cite{Wagatsuma2017} because the injection rate in the present study was two orders of magnitude lower for the slower reaction rate.

 \subsection{Structured cells}

\begin{figure*}
    \centering
     \includegraphics[width=0.96\textwidth]{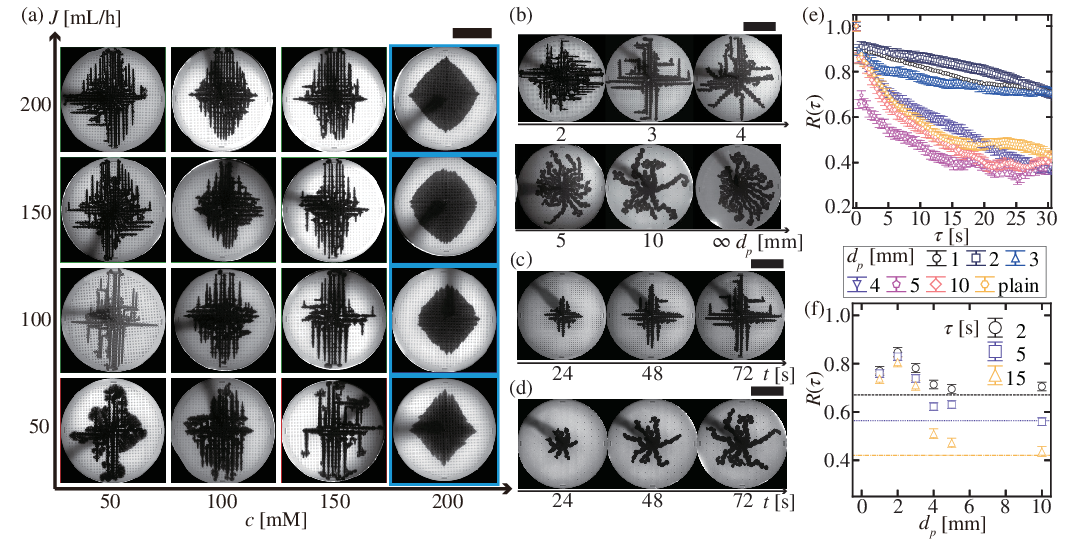}
     \caption{(a) Phase diagram of injection pattern for different $c$ and $J$ in structured cell at $d_p=1$ mm. Blue-shaded and no-shaded regions correspond to a rhombus and filament pattern. The images were taken after 4 mL of inner fluid was injected. 
     (b) Snapshots of the injection pattern after 4 mL of inner solution were injected with different lattice constant $d_p$, where $c$ = 50 mM, $J$ = 150 mL/h. Time development is also shown for $d_p$ =(c) 2 mm and (d)10 mm. Scale bar: 50 mm. (e) Auto-correlation function $R(\tau)$ for different $d_p$. (f) $R(\tau)$ for $\tau$ = 2 s (black circles), 5 s (blue squares), and 15 s (yellow triangles) plotted with $d_p$. $R(\tau)$ without lattice is indicated using black dashed ($\tau$ = 2 s), blue dotted ($\tau$ = 5 s) and yellow dashdotted lines ($\tau$ = 15 s), respectively. Error bars correspond to 99 \% compatible interval.}
      \label{Fig:3}
 \end{figure*}

Experiments with $c$ and $J$ in a plain cell confirmed that the system exhibited filament formation over a wide range of parameter space. To confirm the effect of the structured cell, we injected fluid into a structured cell with regularly arrayed pillars. To observe the typical behavior, the results with the lattice constant of the pillar $d_p = 1$ mm are shown in Fig.~\ref{Fig:3}(a). At a high $c$, instead of a circular front, the front exhibited a rhombus shape. The 4-fold symmetry was due to a square lattice of pillars, while the printing direction of the top plate resulted in the observed anisotropy between the horizontal and vertical directions. For a smaller $c$, we observed filament formation as observed in a plain cell. The filament width was observed to be slightly narrower than that of the plain cell. In addition, the direction of the filament extension was mainly limited to the vertical and horizontal directions, with a slight preference for the vertical direction owing to the printing direction, for the same reason as for the rhombus case. Furthermore, the filaments exhibited minimal meandering. This is a striking effect of the lattice structure, which enhances the transportation of the inner fluid much farther in space.

The experiments were conducted to evaluate the effect of the lattice structure on the filament pattern while the lattice constant $d_p$ was varied as a parameter. The other parameters were fixed at $c$ = 50 mM and $J$ = 150 mL/h, where approximately nine actively extending filaments are expected to coexist simultaneously. Snapshots taken after injecting 4 mL of the inner fluid are shown in Fig.~\ref{Fig:3} (b). For $d_p \leq$ 4 mm, filament growth was affected by the lattice, and filaments extended in a straight manner (Fig.~\ref{Fig:3} (c)) with a filament primarily developed in the horizontal and vertical directions. The filaments exhibited meandering instability at larger values of $d_p$ (Fig.~\ref{Fig:3} (d)) and appeared to be unaffected as in the case without pillars (denoted as $d_p=\infty$ in Fig.~\ref{Fig:3} (b)).

To quantify the stabilization of the extending direction owing to the presence of the pillars, we traced a filament front from the time difference between the images with 2 s time intervals. Images were acquired with $\Delta t=0.4$ s. The director of the filament extension $\bm{n}$ was calculated as $\bm{n}=\bm{v}/|\bm{v}|$. The autocorrelation function of director $R(\tau)$, defined as follows:
\begin{align}
    &R(\tau=\ell\Delta t) \nonumber \\
   & = \frac{1}{\sum_i (N_i-\ell)} \sum_i \sum_{m=0}^{N_i-\ell} \bm{n}_i((m+\ell)\Delta t) \cdot \bm{n}_i(m\Delta t),
\end{align}
were then calculated, where the subscript $i$ denotes $i$-th active filaments whose trajectory was traced for $N_i \Delta t$. The result of $R(\tau)$ is plotted in Fig.~3(e). The persistent direction of the filament growth is enhanced for $d_p$ = 1 -- 3 mm. On the contrary, $R(\tau)$ shows little change in the case of $d_p$ = 4 -- 10 mm. Such an effect of lattice constant $d_p$ can be seen clearly when $R(\tau)$ is plotted against $d_p$, as shown in Fig.~3(f). For $\tau$ = 2 s, the difference is not so evident between the cases with and without lattice, but the difference is seen for $\tau$ = 15 s. Furthermore, the peak in the $R(\tau)$ was observed around $d_p$ = 2 mm, indicating the existence of an appropriate pillar structure. 

The persistent direction of filament growth was enhanced for $d_p$= 1-- 3 mm. In contrast, $R(\tau)$ exhibits minimal change in the case of $d_p$= 4 -- 10 mm. This effect of the lattice constant $d_p$ can be observed clearly when $R(\tau)$ is plotted against $d_p$, as shown in Fig.~\ref{Fig:3} (f). For $\tau$ = 2 s, the difference is not so evident between the cases with and without the lattice; however, the difference is observed for $\tau$=15 s. Furthermore, the peak in the $R(\tau)$ was observed at approximately $d_p$= 2 mm, indicating the existence of an appropriate pillar structure for stabilizing the direction of filament motion.

 \section{Mathematical model}
 \begin{figure}
    \centering
     \includegraphics[width=0.48\textwidth]{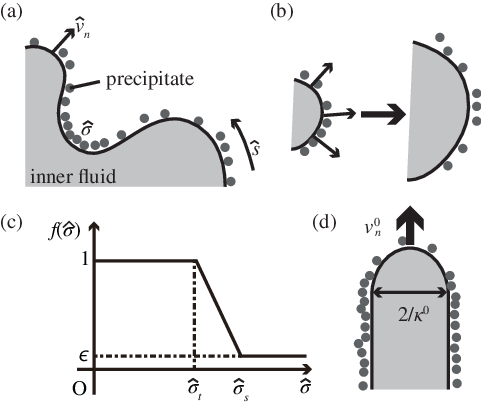}
     \caption{(a) Schematic representation of the boundary dynamics model. The model is based on the movement of a 2D boundary. Outwards normal velocity is represented by $\hat{v}_n$, and the local density of precipitates is represented by $\hat{\sigma}$. $\hat{S}$ indicates a curve of the entire boundary. (b) Decreased $\hat{\sigma}$ owing to curvature effect. (c) The shape of the mobility function $f(\hat{\sigma})$. (d) The simplified situation in which the advancing tip moves with velocity $v_n^0$ and width $2/\kappa^0$.}
     \label{Fig:4}
 \end{figure}
 A model to reproduce the behavior of precipitation-induced filament formation was built, focusing on the relationship with structured cells. Our model is based on the moving boundary model proposed in our previous study~\cite{Wagatsuma2017} as shown in Fig.~\ref{Fig:4}(a). Here, a hat is used to show the variables with physical dimensions. Each segment of the moving boundary is labeled with coordinate $\hat{s}$. The boundary moves in its normal direction with speed $\hat{v}_n(\hat{s})$, while maintaining the local density of the precipitates, $\hat{\sigma}(\hat{s})$. $\hat{v}_n$ and $\hat{\sigma}$ develop with, 
 \begin{align}\label{Eq:Sup1}
  \hat{v}_n=\hat{\xi}f\left(\hat{\sigma}\right) \left(\Delta \hat{p}-\hat{\gamma}\hat{\kappa} \right),
 \end{align}
 and
 \begin{align}\label{Eq:Sup2}
 \frac{\partial \hat{\sigma}}{\partial \hat{t}}=\hat{a} -\hat{\kappa}\hat{\sigma}\hat{v}_n.
 \end{align}
 In Eq.~\eqref{Eq:Sup1},  $\Delta \hat{p}$ represents the pressure difference between the inner and outer fluid. $\hat{\gamma}\hat{\kappa}$ is the Laplace pressure preventing the deformation of the boundary, where $\hat{\gamma}$ is the effective interfacial energy and $\hat{\kappa}$ is the curvature of the boundary defined at $\hat{s}$. In Eq.~\eqref{Eq:Sup2}, $\hat{\sigma}$ increases linearly with time because of precipitation with coefficient $\hat{a}$. The second term represents the change $\hat{\sigma}$ owing to the deformation of the boundary. The unit length in the moving boundary increases to be $1+\hat{\kappa}\hat{v}_n$ with each unit of time; thus, the density of precipitates at the boundary effectively decreases as shown in Fig.~\ref{Fig:4} (b)~\cite{Nakanishi2006}.
 
 In Eq.~\eqref{Eq:Sup1}, $\hat{\xi}f\left(\hat{\sigma}\right)$ represents the mobility of the boundary under the influence of the precipitates. Here, we used the normalized part of mobility $f\left(\hat{\sigma}\right)$ with a piecewise linear function: 
 \begin{align}\label{Eq:Sup3}
     f(\hat{\sigma})=
     \left\{ 
     \begin{matrix}
         1 & (0 \leq \hat{\sigma} \leq \hat{\sigma}_t), \\
         1-(1-\epsilon)\frac{\hat{\sigma} - \hat{\sigma}_t}{\hat{\sigma}_s -\hat{\sigma}_t}& (\hat{\sigma}_t \leq \hat{\sigma} \leq \hat{\sigma}_s), \\
         \epsilon & (\hat{\sigma}_s<\hat{\sigma} ),
     \end{matrix}
     \right.
 \end{align}
 where the function is plotted on the graph shown in Fig.~\ref{Fig:4} (c). This mobility function is intended to have a nonlinear response on the density of the precipitates, where the mobility is unchanged until the density reaches $\hat{\sigma}_t$, decreases linearly until it reaches $\hat{\sigma}_s$, and the boundary almost stops however, with finite mobility $\epsilon$. When $\hat{\sigma}_t=\epsilon=0$, the model is identical to that used in a previous study~\cite{Wagatsuma2017}.
 The inner fluid pressure increases owing to the constant injection and decreases when the moving front advances faster than the injection rate $\hat{J}$. Adopting the effect of finite rigidity of the system $\hat{\Xi}$, $\Delta \hat{p}$ follows,
 \begin{align}\label{Eq:Sup4}
     \frac{\partial \Delta \hat{p}}{\partial \hat{t}}=\hat{\Xi}\left(\frac{\hat{J}}{\hat{h}}-\int_{\hat{S}} \hat{v}_n d \hat{s}\right),
 \end{align}
 where $\hat{h}$ corresponds to the thickness of the cell, and the integral is taken from the entire boundary, $\hat{S}$.
 
 For compact notation, we introduce the units of length, time, and density of aggregate as follows:
 \begin{align}
     \hat{\ell}_c \equiv \left( \frac{\hat{\gamma}}{\hat{\Xi}}\right)^{1/3},  \quad \hat{t}_c\equiv \frac{\hat{\ell}_c^2}{\hat{\xi}\hat{\gamma}}=\left(\frac{1}{\hat{\xi}^3 \hat{\Xi}^2 \hat{\gamma}}\right)^{1/3}, \quad \hat{\sigma}_s. 
 \end{align}
 Then, we use dimensionless variables:
 $t=\hat{t}/\hat{t}_c$,$s=\hat{s}/\hat{\ell}_c$,$v_n=\hat{v}_n\hat{t}_c/\hat{\ell}_c$, $\sigma=\hat{\sigma}/\hat{\sigma}_s$, $\kappa=\hat{\kappa}\hat{\ell}_c$,$\Delta p=\Delta\hat{p} \hat{\ell}_c/\hat{\gamma}$ and dimensionless parameters:
 $\sigma_t=\hat{\sigma}_t/\hat{\sigma}_s$, $a=\hat{a}\hat{t}_c/\hat{\sigma}_s$, $J=\hat{J}\hat{t}_c/(\hat{h}\hat{l}_c^2)$. 
 
 The dimensionless notations of Eqs.~\eqref{Eq:Sup1}-\eqref{Eq:Sup4} are:
 \begin{equation}\label{Eq:Sup5}
     v_n= f(\sigma)(\Delta p - \kappa),
 \end{equation}
 \begin{equation}\label{Eq:Sup6}
     \frac{\partial \sigma}{\partial t}= a-\kappa \sigma v_n,
 \end{equation}
 \begin{align}\label{Eq:Sup7}
     f(\sigma)=
     \left\{ 
         \begin{matrix}
             1 & (0 \leq\sigma \leq \sigma_t), \\
             1-(1-\epsilon)\frac{(\sigma-\sigma_t)}{1 -\sigma_t} & (\sigma_t \leq\sigma \leq 1), \\
             \epsilon & (1<\sigma),
         \end{matrix}
     \right.
 \end{align}
 and
 \begin{align}\label{Eq:Sup8}
     \frac{\partial \Delta p}{\partial t}=J - \int_S v_n ds.
 \end{align}
 The existence of minimal pressure to advance a boundary with a temporally stable density can be predicted using Eqs.~\eqref{Eq:Sup5}-\eqref{Eq:Sup8} with $\epsilon=0$ (see Appendix~\ref{app_steady}). With this condition, a boundary has the steady curvature $\kappa^0$, and normal speed $v_n^0$ determined only by $\sigma_t$ and $a$. If the front of the active filament satisfies this condition (Fig.~\ref{Fig:4}(d)), the extending speed is $v_n^0$, the width is $2\kappa^0$, and the flow rate in the filament is $j^0=2v_n^0 /\kappa^0$. Furthermore, if all the filaments satisfy this condition, the number of active filaments cannot exceed $N_0=J/j_0$. This is a simplified view of the saturation in the number of active filaments. The filament tip could not maintain the above conditions in our numerical calculations, and complex boundary dynamics were observed.

\section{Numerical simulation}\label{app_numerical}
\begin{figure}
   \centering
    \includegraphics[width=0.48\textwidth]{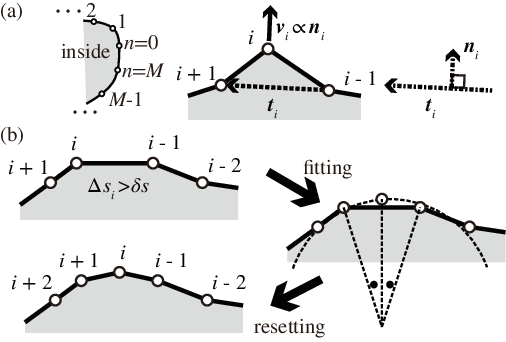}
    \caption{Method for conducting numerical simulations with discrete representative points. (a) A boundary is discretized by representative points denoted by $i$ where $i=0$ to $M(t)$ with tangential and normal vector given by $\bm{t}_i$ and $\bm{n}_i$. (b) Protocol for adding a new representative point when $\Delta s_i > \delta s_i$. A circle is fitted to the position of the four points $j=i-2, \dots i+1$, and a new point is added at the middle point on the circle in angle between $j=i-1, i$.}
    \label{Fig:5}
\end{figure}

Numerical calculation is conducted by modeling moving boundary with representative points denoted by $i$ in 2-dimensional space whose total number is $M(t)$ ($i$= $0 \dots, M(t)$) as shown in Fig.~\ref{Fig:5}(a). Each point has two spatial coordinates and $\sigma$, as $(x_i, y_i, \sigma_i)=(\bm{x}_i, \sigma_i)$. Contour length $\Delta s_i$ is defined by 
\begin{align}
\Delta s_i=\left\{(x_i-x_{i-1})^{2} +(y_i-y_{i-1})^{2}\right\}^{1/2}=|\bm{x}_{i-1}-\bm{x}_i|.
\end{align}
The coordinate $s_i$ is given by $s_i =\sum_{j=0}^{i} \Delta s_j$. As shown in Fig.~\ref{Fig:5}(a), the outward normal vector $\bm{n}_i$ for the point $i$ is defined by $\bm{n}_i \cdot \bm{t}_i=0$, $\left| \bm{n}_i \right|=1$, and 
\begin{align}
\bm{t}_i=(x_{i+1}-x_{i-1}, y_{i+1}-y_{i-1}).
\end{align}
The curvature $\kappa_i$ was calculated by
\begin{align}
\kappa_i = \left\{ \left(\frac{d^2 x}{d s^2}\right)_i  \left(\frac{d y}{d s}\right)_i-\left(\frac{d^2 y}{d s^2}\right)_i  \left(\frac{d s}{d s}\right)_i \right\} \cdot \frac{1}{\left| \left(\frac{d \bm{x}}{d s} \right)_i  \right|^3}, 
\end{align}
where
\begin{align}
\left(\frac{d \bm{x}}{d s} \right)_i &= \frac{\bm{x}_{i+1}-\bm{x}_{i-1}}{s_{i+1}-s_{i-1}}= \frac{\bm{x}_{i+1}-\bm{x}_{i-1}}{\Delta s_{i+1}+\Delta s_{i}}\\
\left(\frac{d^2 \bm{x}}{d s^2} \right)_i &= \nonumber \\
 &\left(\frac{\bm{x}_{i+1}-\bm{x}_{i}}{s_{i+1}-s_{i}}- \frac{\bm{x}_{i}-\bm{x}_{i-1}}{s_{i}-s_{i-1}}\right)\cdot \frac{2}{s_{i+1}-s_{i-1}}.
\end{align}
In each time step, $\bm{x}_{i}$, $\sigma_i$ and $\Delta p$ is updated to be $\bm{x}_{i}^N$, $\sigma_i^N$ and $\Delta p^N$ by following
\begin{equation}\label{sim:eq0-nd}
	\bm{x}_{i}^N= \bm{x}_{i}+\left(v_n\right)_i \bm{n}_{i} \Delta t,
\end{equation}
\begin{equation}\label{sim:eq1-nd}
	\sigma_i^N= \sigma_i+\left(a-\kappa_i \sigma_i \left(v_n\right)_i \right) \Delta t,
\end{equation}
\begin{equation}\label{sim:eq2-nd}
	\left(v_n\right)_i=f( \sigma_i)(\Delta p - \kappa_i),
\end{equation}
and
\begin{equation}\label{sim:eq3-nd}
	\Delta p^N=\Delta p + \left(J  - \sum_i \Delta s_i \left(v_n\right)_i\right) \Delta t.
\end{equation}

The representative points are updated by the following rule. A new representative point is added when $\Delta s_i > \delta s$ as in Fig.~\ref{Fig:5}(b). A circle is fitted to four representing points $\bm{x}_{i-2}$, $\bm{x}_{i-1}, \bm{x}_{i}, \bm{x}_{i+1}$ (see see Appendix~\ref{app_circlefit}), and the angle between  
$\bm{x}_{i-1}, \bm{x}_{i}$ at the center of the fitted circle is evenly divided into 1/2 by the newly added point, and the point is renumbered as $\bm{x}^N_{i-2}, \bm{x}^N_{i-1}, \bm{x}^N_{i}, \bm{x}^N_{i+1}, \bm{x}^N_{i+2}$, while $\bm{x}^N_{i-2}=\bm{x}_{i-2}$, $\bm{x}^N_{i-1}=\bm{x}_{i-1}$, $\bm{x}^N_{i+1}=\bm{x}_{i}$, $\bm{x}^N_{i+2}=\bm{x}_{i+1}$. $\bm{x}^N_{i}$ is the position of the newly added point. This refinement of the representative points was conducted at the refresh rate of $\Delta t_r$. The total number of representative points $M(t)$ increases with time.

When any representative points $k$ with $|i-k| \geq 2$ comes close to the point $i$, that is $|\bm{x}_{i}-\bm{x}_{k}| < \delta s$, both point $k$ and $i$ stop their motions. This way, we avoid the boundary's overlap. Furthermore, we prepared pillars with square lattice configuration, while the lattice constant is set as $d_p$, and the radius of the pillar is denoted as $r_p$. When representative points enter the pillar region, the representative points stop motion.

The temporal integration was conducted with simple Euler methods. The initial condition was a circle with radius 1, and the representative point was set evenly with distance $\delta s/2$. The size of the circle must be finite to overcome the Laplace pressure, which shrinks the boundary. A uniform noise was given in the initial value $\sigma_i$ in 0.5 to 1. The initial value of $\Delta p=0$. $r_p$ is set small and fixed as 0.25. We used $\Delta t=0.0001$, $\delta s=0.1$, and the refresh rate of the representative points to be $\Delta t_r = \delta s/100=0.001$. We used $\epsilon = 0.01$ as a small parameter. We used $\sigma_t = 0.9$ unless stated explicitly.
The remaining parameters are $a$, $J$, and $d_p$. 


\begin{figure}
    \centering
     \includegraphics[width=0.48\textwidth]{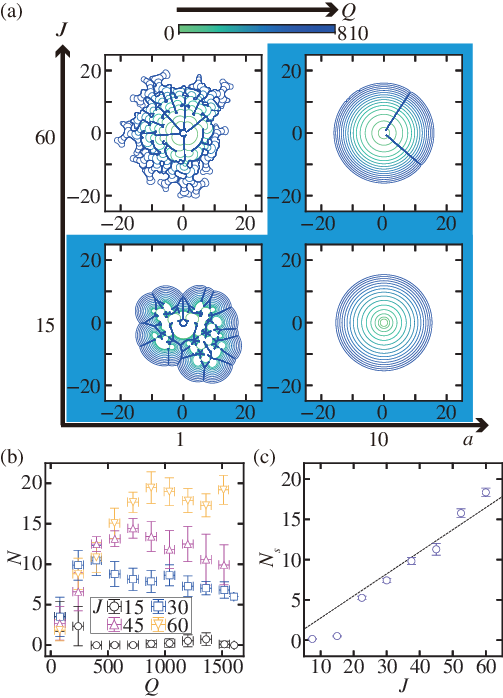}
     \caption{
     Result of numerical simulations. (a) Phase diagram of injection patterns for different $a$ and $J$ without pillars. Lines correspond to the boundary position at a given amount of injected fluid $Q = Jt$ denoted by colors given at the top bars. The final images were taken at $Q$ = 810. (b)Time course of the number of active filaments $N$ for $a=1$ without the presence of the pillars. The inset denotes the symbol for each $J$. The error bar corresponds to the standard deviation. (c) The saturated number of active filaments $N_s$ for $a=1$. The data averaged over $Q=1000$ to $1500$. The dashed line represents the linear fitting.}
     \label{Fig:6}
 \end{figure}


\begin{figure}
    \centering
     \includegraphics[width=0.48\textwidth]{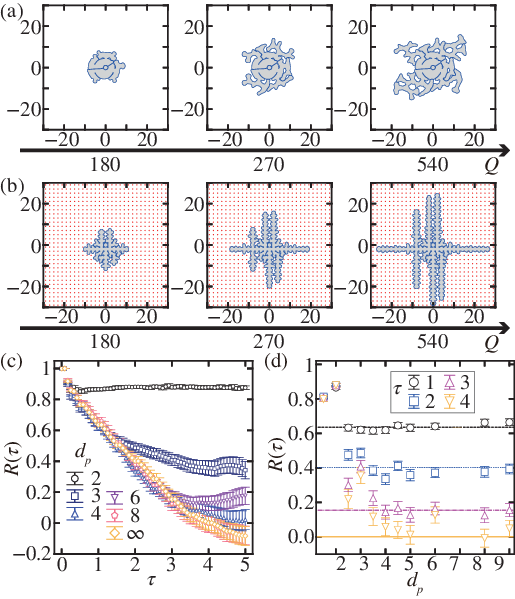}
     \caption{
     Result of numerical simulations. (a) without lattice structure and (b) with lattice structure $d_p= 2$. (c) Autocorrelation function $R(\tau)$ for different $d_p$. The inset denotes the symbol color for each $d_p$. (d) $R(\tau)$ is plotted for each $d_p$. $R(\tau)$ without lattice is indicated by the lines for each $\tau$, and their colors correspond to the symbol colors of the plot of $R(\tau)$.  Error bars correspond to 99 \% compatible intervals.}
     \label{Fig:7}
 \end{figure}



The results of the numerical calculations are shown in Fig.~\ref{Fig:6}. We reproduced the phase diagram by changing the reaction rate $a$ and injection rate $J$, where filament patterns appeared at low $a$, corresponding to low $c$ in the experiment, and high $J$ as shown in Fig.~\ref{Fig:6}(a). At $J = 15$ and $a = 1$, we observed a filament pattern initially, then turned into a homogeneous extension of a boundary. Such a transition was also observed in the experiment.
Fig.~\ref{Fig:6}(b) shows the number of active filaments saturated with time. Here, we plotted $N$ against the amount of injected fluid, $Q = Jt$. $N$ increased initially with time and almost reached a maximum steady value. However, $N$ exhibited a slight decay later. This is attributed to the finite mobility $\epsilon$ for $\sigma > 1$. We observed that such a decrease in the active filament number was eliminated in the case with $\epsilon=0$. In such a case, the boundary motion can be terminated completely if the injection rate is not large enough as all boundaries reach $\sigma>1$. As we never observed the complete termination of the boundary motion during the injection process, we used non-zero $\epsilon$ in our mathematical model. Fig.~\ref{Fig:6}(c) shows the saturated number of active filaments, which was proportional to the injection rate. The dashed line corresponds to the fitting with $N_s= \alpha J$, whereas $\alpha=6.35 \times 10^{-2}$ $\pm 1.3 \times 10^{-3}$ h/mL.

Furthermore, the model showed extensive meandering in the absence of pillars as shown in Fig.~\ref{Fig:7}(a), while the model reproduced the suppression of meandering motion by a regular array of pillars as shown in Fig.~\ref{Fig:7}(b). These extensive meandering and suppression of filament were confirmed quantitatively by the autocorrelation function of the filament motion direction as shown in Fig.~\ref{Fig:7}(c). The suppression of meandering was most enhanced at $d_p = 2$, which indicates the appropriate pillar structure as well as the experiment.

\begin{figure}
	\centering
	 \includegraphics[width=0.48\textwidth]{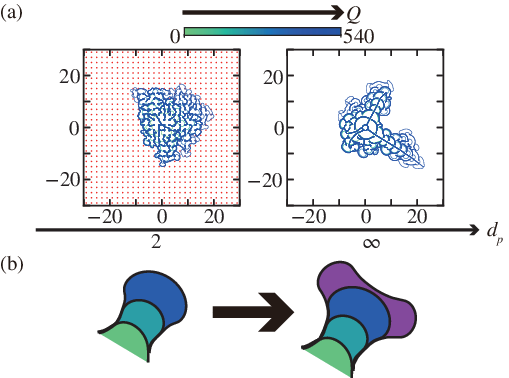}
	 \caption{
		(a) Effect of $\sigma_t$, where $\sigma_t$ was set to 0, while maintaining $a = 1$, $J = 30$. Colored lines correspond to the boundary and indicate the dynamics based on the color bar. The final images were obtained at $Q = 540$. 
        (b) Typical shape of filament tip with $\sigma_t=0.9$, front flattened. This shape induces tip splitting as well as the meandering motion of the filament.}
	 \label{Fig:8}
 \end{figure}

In our numerical simulation, we used $\sigma_t=0.9$. In the case with $\sigma_t$ is not large enough, the meandering of the front motion was significant so that the front does not show straight motion even in the presence of pillars as shown in Fig.~\ref{Fig:8}(a). To reproduce the observed boundary dynamics, we need to have the mobility function of precipitate density to be nonlinear step-like as shown in Fig.~\ref{Fig:4}(c) with finite mobility for high $\sigma$.

\section{Discussion}
Here, we discuss the dynamics of filament motion based on the mathematical model. One key feature observed is tip splitting. Extending fronts increased their number at the beginning of the numerical simulation by tip-splitting. However, once the number of moving fronts increases, the value of the integral in Eq.~\eqref{Eq:Sup8} becomes larger than $J$, $\Delta p$ starts to decrease. Remember that the front speed decreases as in Eq.~\eqref{Eq:Sup5} for smaller $\Delta p$, and the decreased front speed tends to increase the local density of precipitates $\sigma$ owing to the term $-\kappa \sigma v_n$, further reducing mobility. Hence, there is an autocatalytic process to lower the boundary speed triggered by the decrease in $\Delta p$. Through this process, the split fronts stopped moving. This suspension of tip splitting occurs at the late stage of numerical simulations, where the number of active filaments is saturated. Instead, suspended tip-splitting is observed as the meandering behavior of the filaments. Thus, the tip-splitting behavior of the filament is crucial to the filament motion. 

The details of tip splitting can be explained by the following simple argument as shown in Fig.~\ref{Fig:8}(b). We use the observed fact from the numerical simulation that a small $\sigma_t$ leads to the failure of the straight motion with pillars (Fig.~\ref{Fig:8}(a)) and take $\sigma_t \sim 1$ in the following argument. Under this condition, the mobility of the boundary is independent of $\sigma$ and suddenly approaches zero when $\sigma=1$. Subsequently, the extending fronts have a flat curvature due to the larger normal velocity $v_n$ and a smaller $\kappa$. The flat boundary quickly accumulates precipitates, that is, $\sigma$ increases rapidly owing to the absence of $-\kappa \sigma v_n$ in Eq.~\eqref{Eq:Sup5}. Once the leading flat boundary of the extending filament stops, the side parts with a higher curvature start to extend because of the increased pressure caused by the decreased area increase over time. Consequently, the two boundaries begin to extend from the side as shown in Fig.~\ref{Fig:8}(b), that is, the tip-splitting behavior.

Tip splitting of the active filament is suppressed in the case of high pillar density by imposing curvature on the active front. Here, we illustrate the effect of the pillar to suppress tip splitting by a sufficiently small $d_p$. The dynamics of $\sigma$ (Eq.~\eqref{Eq:Sup5}) can be discretized in time as, 
$\delta \sigma= \delta t (a-\kappa \sigma v_n)$,
where the discretization was performed for a front sweep of one unit cell of the pillars. We can take the curvature of the front $\kappa \sim 2/d_p$ imposed by the pillars. Other values are almost constant as $\sigma \sim \sigma^0 \sim 1$ and $v_n \sim v_n^0 \sim \sqrt{a}$ because of the small $\delta t$, where $\sigma^0$ and $v_n^0$ represents the values of steadily extending filament with minimal pressure (see Appendix A). Subsequently, we have the condition for the steadily moving front affected by the presence of pillars $\delta \sigma =0$ as,
$a-2\sqrt{a}/d_p=0$.
The moving front did not stop under $d_p = 2/\sqrt{a}$, and meandering did not occur.

Our numerical simulation indicates the relevance of the nonlinear step-like response of the mobility of the boundary on the density of the precipitate density. We observed that the mobility function can be altered to be a smooth step-like function, as well. Such a step-like nonlinear response of the mobility can be justified if we consider the change in the mobility owing to the appearance of solid-solid friction once the boundary is fully covered by the precipitates. Unless it is covered completely by the precipitates, the boundary is insensitive to the density and mobile as a fresh one. Such a steep change in the mobility is necessary to have $\sigma_t$ large and to have a finite time to have straight motion of the boundary. 

Finally, we discuss the transition from the filament pattern to the circular pattern, as shown in Fig.~\ref{Fig:6}(a). This is because of the small but constant and finite boundary mobility for $\sigma>1$. In this case, the mobility did not depend on $\sigma$, and the boundary extends while keeping the curvature as small as possible owing to the effect of the Laplace pressure. In this way, the front shows a circular pattern. Physically, the model suggests that the boundaries' mobility becomes insensitive to the precipitate density when the density is high enough.

\section{Conclusion}
In this study, we observed the reaction-injection pattern with precipitation using plain and structured cells. We used the combination of CTAB and NaSal aqueous solutions. We found the transition from the circular pattern to the filament pattern with a plain cell (Fig.~\ref{Fig:2}(a)). In the filament pattern, the active filaments exhibited meandering during the extension, a halt of their motion and split of the front. We confirmed that the number of active filaments $N$ saturated to be $N_s$ (Fig.~\ref{Fig:2}(f)), which depended linearly on the injection rate of inner fluid $J$ (Fig.~\ref{Fig:2}(e)). These observations are consistent with previous reports~\cite{Podgorski2007, Wagatsuma2017}.

We found the transition from the rhombus pattern to the filament pattern (Fig.~\ref{Fig:3}(a)) with a structured cell(Fig.~\ref{Fig:1}(a)). In the parameter region corresponding to the circular pattern, the rhombus pattern with 4-fold symmetry was observed. In the case of the filament pattern, each filament showed a straight extension, which offers a clear contrast with the meandering motion in a plain cell. By changing the lattice constant of the pillar $d_p$, we confirmed the suppression of the meandering motion was the most distinctive when $d_p =$ 2 mm (Fig.~\ref{Fig:3}(f)). 

We built a mathematical model based on the dynamics of boundaries. The model incorporated the effect of the Laplace pressure that forces the boundary to take a flat shape. In addition, the mobility of the boundary that nonlinearly depends on the local density of precipitates $\sigma$ was considered. Our model successfully reproduced the transition from the circular pattern to the filament pattern by the injection rate $J$ and reaction rate $a$ (Fig.~\ref{Fig:6}(a)) and linear dependence of $N_s$ on $J$ (Fig.~\ref{Fig:6}(e)). Further, we reproduced the effect of pillars on the filament pattern (Fig.~\ref{Fig:7}(a) and (b)). 

Significantly, the suppression of the meandering behavior of the filaments is reproduced by the effect of pillars with appropriate lattice constant $d_p$ (Fig.~\ref{Fig:7}(d)). We explained the meandering mechanism resulting from tip splitting (Fig.~\ref{Fig:8}(b)) and estimated the appropriate lattice constant, $d_p$, to suppress meandering. This estimation is consistent with the value determined in the mathematical simulation. We also explained the circular extension of the boundary owing to the homogeneous small mobility due to high $a$. We further discussed that the meandering motion of the filament is caused by the nonlinear dependence of the mobility of the boundary on the local density of the precipitate.

Our study revealed the relevance of the precipitate formation process in the injection of fluid into a confined geometry. Especially the transition from the circular injection front to the filament pattern by the injection rate may allow the injected fluid to be transported further than expected. The suppression of the meandering motion of the filament may enhance the transportation of the injected fluid. 
The transportation of fluid has a large impact on geological events and civil engineering procedures, such as injection-induced earthquakes and chemical grouting.
Our experimental and theoretical findings will shed light on the relevance of the coupling between the flow and the precipitation with the nonlinear mobility of the boundaries in these processes.

\subsection*{Acknowledgment}
  This work was supported by JSPS KAKENHI Grant JP16H06478, JP21H00409, and JP21H01004 to YS. and by JSPS and PAN under the Japan-Poland Research Cooperative Program ``Spatio-temporal patterns of elements driven by self-generated, geometrically constrained flows'', the Cooperative Research of ``Network Joint Research Center for Materials and Devices'' with Hokkaido University (Nos.~20181048) to YS. This work was also supported by  JST, the establishment of university fellowships towards the creation of science technology innovation (Grant No. JPMJFS2144) to KO. 

 S. Tanaka and K. Otoguro contributed equally to this work.

\appendix
\setcounter{section}{0}
\renewcommand{\thesection}{\Alph{section}}

\section{Minimal pressure for moving boundary}\label{app_steady}
\begin{figure}
   \centering
    \includegraphics[width=0.48\textwidth]{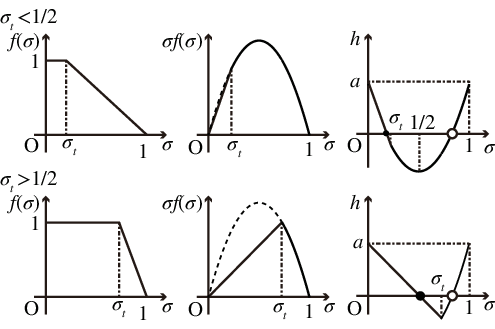}
    \caption{Shape of normalized mobility $f(\sigma)$, $\sigma f(\sigma)$ and $h(a, \sigma)$ for $\sigma_t<1/2$ and $\sigma_t>1/2$  Here we plotted by assuming $\kappa(\Delta p-\kappa)>0$.}
    \label{Fig:9}
\end{figure}

Here, we discuss the dynamics of interfacial motion of our mathematical model, which is a modification of our previous one shown in~\cite{Wagatsuma2017}. Here, we assume $\epsilon=0$ for simplicity. At first, let us assume $\Delta p$, and $\kappa$ constant. From
\begin{equation}\label{app1}
	\frac{d \sigma}{d t}= a-\kappa \sigma v_n,
\end{equation}
\begin{equation}\label{app2}
	v_n= f(\sigma)(\Delta p - \kappa),
\end{equation} with
\begin{align}\label{app3}
	f(\sigma)=
	\left\{ 
	\begin{matrix}
	    1 & (0 \leq\sigma \leq \sigma_t), \\
		1- (\sigma-\sigma_t)/(1 -\sigma_t) & (\sigma_t \leq\sigma \leq 1), \\
		0 &  (1<\sigma ),
	\end{matrix}
	\right.
\end{align}
we obtain
\begin{equation}\label{app4}
	\frac{d\sigma}{d t}= a-\kappa \sigma f(\sigma)(\Delta p - \kappa)=h(a,\sigma).
\end{equation} 
We sketch the graph of $f(\sigma)$,  $\sigma f(\sigma)$ and $h(a, \sigma)$ in Fig.~\ref{Fig:9}. We find the stable fixed point appear only when $h(a,1/2)<0$ for $\sigma_t < 1/2$ and $h(a,\sigma_t) <0$ for $\sigma_t > 1/2$. Otherwise, $\sigma$ increases with time; eventually, the moving boundary stops due to zero mobility in the case of $\sigma \geq 1$. 
To have a steadily moving boundary, thus, 
\begin{align}\label{app4-1}
\sigma^0=\left\{
\begin{array}{c c}
1/2 & (\sigma_t<1/2), \\
\sigma_t & (\sigma_t\geq 1/2),
\end{array}
\right.
\end{align}
and from $h(a, \sigma^0)\leq0$,
\begin{align}\label{app5}
\kappa (\Delta p - \kappa) \geq \Pi,  
\end{align}
where
\begin{align}\label{app6}
\Pi=\left\{
\begin{array}{c c}
4a(1-\sigma_t) & (\sigma_t<1/2), \\
a/\sigma_t & (\sigma_t\geq 1/2).
\end{array}
\right.
\end{align}

To extract the minimum pressure to have a moving boundary, we should note that the left-hand side of Eq.\eqref{app5}, $\kappa (\Delta p - \kappa)$, has a maximum when $\kappa$ is varied, and the maximum is $(\Delta p)^2/4$ at $\kappa = \Delta p/2$.
One can see that $(\Delta p)^2 \geq 4\Pi$ is the condition to specify the minimum pressure to have a steadily moving boundary, and the minimum pressure $\Delta p^0$ is given by
\begin{align}\label{app6-2}
\Delta p^0 =\left\{
\begin{array}{c c}
4\sqrt{a(1-\sigma_t)} & (\sigma_t<1/2), \\
2\sqrt{a/\sigma_t} & (\sigma_t\geq 1/2).
\end{array}
\right.
\end{align}
Here, the possible curvature to have a steadily moving boundary is given by $\kappa^0=\Delta p_0/2$, and
\begin{align}\label{app6-3}
\kappa^0 =\left\{
\begin{array}{c c}
2\sqrt{a(1-\sigma_t)} & (\sigma_t<1/2), \\
\sqrt{a/\sigma_t} & (\sigma_t\geq 1/2).
\end{array}
\right.
\end{align}
Further, such condition imposes the speed of boundary $v^0_{n}$ is given by 
\begin{align}\label{app6-4}
v_n^0 =\left\{
\begin{array}{c c}
\sqrt{a/(1-\sigma_t)} & (\sigma_t<1/2), \\
\sqrt{a/\sigma_t} & (\sigma_t\geq 1/2).
\end{array}
\right.
\end{align}
Care should be taken that in the condition with the minimum pressure $\Delta p^0$, the curvature $\kappa^0$ and the velocity $v^0_{n}$ is determined only by $\sigma_t$.

These analyses are identical to the one done in~\cite{Wagatsuma2017}. As shown in Fig.~\ref{Fig:4}(d), if we assume the tip part of an extending filament satisfies the above-mentioned steady state with the constant curvature $\kappa^0$, the width$2/\kappa^0$ and the speed $v^0_{n}$, the flow rate of the single filament is $j^0=2v_n^0/\kappa^0$, where
\begin{align}\label{app7}
j^0_{n}=\left\{
\begin{array}{c c}
1/(1-\sigma_t) & (\sigma_t<1/2), \\
2 & (\sigma_t\geq 1/2).
\end{array}
\right.
\end{align}
The number of active filaments, thus, saturates at $N_0=J/j_0$.


\section{Circle fit}\label{app_circlefit}
\begin{figure}
  \includegraphics[width=0.48\textwidth]{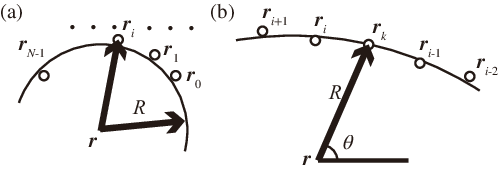}
\caption{(a) Schematic illustration of circle fit. (b) Schematic illustration of interpolation of a new point while keeping the curvature fixed.}
\label{Fig:10}
\end{figure}

Here we show a method to fit $N$ points $(x_i, y_i)$ with $i \in [0,N-1]$ to a circle whose center point and radius are $\bm{r}=(x, y)$ and $R$, respectively (Fig.~\ref{Fig:10}(a)).
As a merit function, we use $f=\sum_i \left(r^2_i-R^2\right)^2$, where $r_i= \left\{\left(x_i-x\right)^2+\left(y_i-y\right)^2 \right\}^{1/2}.$ Taking derivative with respect to $x$, $y$, and $R$, we have
\begin{equation}
\Par{f}{x}=4\sum_i(r^2_i-R^2)(x-x_i),
\end{equation}
\begin{equation}
\Par{f}{y}=4\sum_i(r^2_i-R^2)(y-y_i),
\end{equation}
and
\begin{equation}
\Par{f}{R}=-4R\sum_i(r^2_i-R^2).
\end{equation}
(Note that $\Par{r_i}{x}=\frac{x-x_i}{r_i}$ and hence $\Par{r^2_i}{x}=2(x-x_i)$.) For the condition of minima, $\Par{f}{x}=\Par{f}{y}=\Par{f}{R}=0$, we have,
\begin{equation}\label{xres}
\sum_i(r^2_i-R^2)(x-x_i)=0
\end{equation}
\begin{equation}\label{yres}
\sum_i(r^2_i-R^2)(y-y_i)=0
\end{equation}
and
\begin{equation}\label{Rres}
\sum_i(r^2_i-R^2)=0
\end{equation}

In the following, we use notation  $\langle \cdots \rangle$ as average of $\cdots$ with respect to $i$, that is $\langle \cdots \rangle = (\sum_i \cdots)/\sum_i$. From the condition Eq. (\ref{Rres}), we have $R^2=\Mean{r^2_i}$. Furthermore, we can obtain,
\begin{align}\label{Req}
R^2=&\Mean{r^2_i}=\frac{1}{n}\sum_i r^2_i  \nonumber \\
= & \Mean{x^2_i} +\Mean{y^2_i} -2 x\Mean{x_i}-2 y\Mean{y_i}+ x^2 +y^2 .
\end{align}

Then, we have $\sum_i x_i r^2_i = R^2 \sum_i x_i$, from Eq. (\ref{xres}) with the condition Eq. (\ref{Rres}), leading to
\begin{align}
\Mean{x^3_i}&+\Mean{x_i y^2_i} -2x\Mean{x^2_i}-2y\Mean{x_i y_i} \nonumber \\
&= \Mean{x_i}\left(\Mean{x^2_i} +\Mean{y^2_i} -2 x\Mean{x_i}-2 y\Mean{y_i} \right),
\end{align}
which is
\begin{align}
(\Mean{x^2_i}-&\Mean{x_i}^2)x+(\Mean{x_i y_i}-\Mean{x_i}\Mean{y_i})y \nonumber \\
&=\frac{1}{2}\left(\Mean{x^3_i}+\Mean{x_i y^2_i}-\Mean{x_i}\Mean{x^2_i}-\Mean{x_i}\Mean{y^2_i}\right)
\end{align}
From symmetry, Eq. (\ref{yres}) and  Eq. (\ref{Rres}) leads to,
\begin{align}
(\Mean{x_i y_i}-&\Mean{x_i}\Mean{y_i})x+(\Mean{y^2_i}-\Mean{y_i}^2)y \nonumber \\
&=\frac{1}{2}\left(\Mean{y^3_i}+\Mean{x^2_i y_i}-\Mean{y_i}\Mean{y^2_i}-\Mean{x^2_i}\Mean{y_i}\right)
\end{align}
Finally, we have $x=\phi(x_i,y_i)/\xi$,$y=\phi(y_i,x_i)/\xi$ where
\begin{align}
\phi(x_i,y_i)&=\left(\Mean{x^3_i} +\Mean{x_i y^2_i}\right) \left( \Mean{y^2_i} - \Mean{y_i}^2 \right) \nonumber \\
&-\left( \Mean{y_i^3}+\Mean{x^2_i y_i}\right) \left(\Mean{x_i y_i}-\Mean{x_i}\Mean{y_i} \right) \nonumber \\
&+ \left( \Mean{y_i}\Mean{x_i y_i}-\Mean{x_i} \Mean{y_i^2}\right) \left(\Mean{x_i^2}+\Mean{y_i^2} \right),
\end{align}
and
\begin{align}
\xi=2\left\{(\Mean{x^2_i}-\Mean{x_i}^2)(\Mean{y^2_i}-\Mean{y_i}^2) \right.\nonumber \\
\left.-\left(\Mean{x_i y_i}-\Mean{x_i}\Mean{y_i} \right)^2\right\}.
\end{align}
$R$ is obtained from Eq.~\eqref{Req} with obtained $x$ and $y$.

Using the above fitting with a circle, we interpolate a point $\bm{r}_k=(x_k, y_k)$ from 4 original data points $\bm{r}_j=(x_j, y_j)$ ($j=i-2$ to $i+1$). A new point is inserted between $\bm{r}_{i-1}$ and $\bm{r}_{i}$ (Fig.~\ref{Fig:10}(b)).
The center of the circle $\bm{r}=(x,y)$, as well as the radius $R$ is obtained by fitting four original data points $\bm{r}_j=(x_j, y_j)$ ($j=i-2$ to $i+1$) with a circle. 
Then, 
\begin{equation}
\theta_{i-1} = \arctan\left(\frac{y_{i-1}-y}{x_{i-1}-x}\right), \,\,\, \theta_2=\arctan \left( \frac{y_{i}-y}{x_{i}-x}\right)
\end{equation}
are obtained, and $\theta$ is defined by $\theta=(\theta_{i-1}+\theta_{i})/2$
From a formula of trigonometric functions, 
\begin{align}
\tan \theta &=\tan \frac{\theta_{i-1}+\theta_{i}}{2} \nonumber \\
 &= \frac{2\sin \frac{\theta_{i-1}+\theta_{i}}{2} \cos \frac{\theta_{i-1}-\theta_{i}}{2}}{2\cos \frac{\theta_{i-1}+\theta_{i}}{2} \cos \frac{\theta_{i-1}-\theta_{i}}{2}}=\frac{\sin \theta_{i-1}+\sin \theta_{i}}{\cos \theta_{i-1}+\cos \theta_{i}},
\end{align}
and
\begin{align}
\theta &= \arctan\left(\frac{\frac{y_{i-1}-y}{r_{i-1}}+\frac{y_{i}-y}{r_{i}}}{\frac{x_{i-1}-x}{r_{i-1}}+\frac{x_{i}-x}{r_{i}}}\right) \nonumber \\
&= \arctan\left(\frac{r_{i} (y_{i-1}-y)+r_{i-1}(y_{i}-y)}{r_{i} (x_{i-1}-x)+r_{i-1}(x_{i} -x)}\right)
\end{align}
Then we have $x_k=R\cos\theta+x$, and $y_k=R\sin\theta+y$.

\end{document}